\documentclass[11pt]{article} 
\usepackage{graphicx,authblk,amssymb,amsfonts,amsmath,geometry}
\usepackage{graphicx,amssymb,amsfonts,amsmath,geometry}
\usepackage{dcolumn}
\usepackage{bm,mathtools}
\usepackage{cancel}
\usepackage{soul}
\usepackage[normalem]{ulem}
\usepackage{color}
\usepackage[utf8]{inputenc}
\usepackage[T1]{fontenc}
\usepackage{mathptmx}
\usepackage{physics}

\begin{document}  

\title{Oscillatory instability and stability of stationary solutions in the parametrically driven, damped nonlinear Schr\"odinger equation}

\author[1]{Fernando Carreño-Navas\thanks{E-mail: fercarnav@gmail.com, ran@us.es, niurka@us.es}}
\affil[1]{IMUS, Universidad de Sevilla, 41012, Sevilla, Spain}
\author[2]{Renato Alvarez-Nodarse}
\affil[2]{IMUS and Departamento de An\'alisis Matem\'atico, Universidad deSevilla, c/Tarfia s/n, 41012, Sevilla, Spain}
\author[3]{Niurka R.\ Quintero}
\affil[3]{Departamento de F\'\i sica Aplicada I, Escuela T\'ecnica Superior de Ingenier\'\i a  
Inform\'atica, Universidad de Sevilla, Avda Reina  Mercedes s/n, 41012 Sevilla, Spain}

\date{\today}
\maketitle
 
\begin{abstract}
 We found two stationary solutions of the 
parametrically driven, damped nonlinear Schrö\-dinger equation with 
a nonlinear term proportional to $|\psi(x,t)|^{2 \kappa} \psi(x,t)$ 
for positive values of $\kappa$. By linearizing the equation around these exact solutions, we derived the corresponding Sturm-Liouville problem. 
Our analysis reveals that one of the stationary solutions is unstable, while the stability of the other solution depends on the amplitude of the parametric force, the 
damping coefficient, and the nonlinearity parameter  $\kappa$.  
An exceptional change of variables facilitates the computation of the stability diagram  through numerical solutions of the eigenvalue problem as a specific parameter $\varepsilon$ varies within a bounded interval. 
For $\kappa <2$ , an {\it oscillatory instability}  is predicted analytically and confirmed numerically. 
Our principal result establishes that for $\kappa \ge 2$,  there exists a critical value of $\varepsilon$ 
beyond which the unstable soliton becomes stable, exhibiting {\it  oscillatory stability}.
\end{abstract}

\section{Introduction} \label{sec1}

The nonlinear Schr\"odinger (NLS) soliton under parametric excitation captures the dynamics of  small-amplitude breathers in the easy-plane ferromagnet and the long Josephson junction under the influence of the parametric pumping and dissipation  \cite{barashenkov:1991,bondila:1995}.   
Additionally, it accounts for various phenomena, including
 the repulsive behavior of the double-solitons observed in an oscillating water channel of finite length 
\cite{wang:1997}, the excited surface-waves solitons reported in Ref. \cite{wang:1996,wang:1998}, and the non-propagating hydrodynamic soliton under confinement \cite{gordillo:2014}.

Constant and spatio-temporal parametric forces lead to moving NLS solitary waves which break the integrability of the NLS equation. 
In the former case, stable non-propagating and moving solitons coexist at low driving strengths, while strongly forced solitons are stable only at high speeds 
\cite{barashenkov:2001}.
For time and space dependency of the force, the soliton can oscillate around a certain position, and two empirical stability criteria, based on the collective coordinate approach, have been applied to determine the regions of stability \cite{mertens:2020}.

 Time-dependent parametric pumping in the NLS equation plays a crucial role in various phenomena. It facilitates the creation and annihilation of solitons \cite{wang:2001} and supports the existence of soliton bound states \cite{barashenkov:1999,barashenkov:2011,bogdan:2022}. Additionally, it models small-amplitude breathers in parametrically driven, damped systems such as the Landau-Lifshitz and sine-Gordon equations \cite{bondila:1995}, stabilizing the damped solitons \cite{miles:1984,barashenkov:1991,barashenkov:2001}.
Unlike the nonlinear Klein-Gordon equations \cite{scott:2003,dauxois:2010,sanchez:1998}, where topological solitary waves exhibit robustness in the presence of damping, envelope solitons in the nonlinear Schrödinger equation are damped and eventually fade away under similar conditions. However, the introduction of an additional parametric force, $r \,e^{2 i  t} \Psi^{\star}(x,t)$, with amplitude $r$, can compensate the dissipative term, $i \,\rho \,\Psi(x,t)$, where $\rho$ is the dissipation coefficient. This parametric driving stabilizes stationary solutions of the parametrically driven, damped nonlinear Schrödinger equation
\begin{equation}
\label{p1kappa}
i \Psi_t+\Psi_{xx}+2 |\Psi|^{2\,\kappa} \Psi = r e^{2 i  t} \Psi^{\star}-i \rho \Psi.
\end{equation}
 The purpose of this study is to investigate this equation.  

Equation \eqref{p1kappa} for $\kappa=1$ has been derived from the Landau-Lifshitz 
equation, which describes the magnetization dynamics of ferromagnetic materials \cite{barashenkov:1991}. 
In this model, the azimuthal angle of the unit vector of magnetization $\vec{m}$ is governed by the driven sine-Gordon equation, while  
a linear combination of the $y$ and $z$ components of $\vec{m}$ is directly related to the solution  $\Psi(x,t)$ of Eq.\ \eqref{p1kappa}. 
The NLS equation has been extended for $\kappa \ne 1$ (see \cite{sulem:1999} and reference therein). This generalization corresponds to a non-Kerr law nonlinearity, where the nonlinear response of the index of refraction deviates from the standard square 
dependence of the electric field  
 \cite{biswas:2002}. For specific case of $\kappa=2$, the generalized NLS equation describes the collapse of a plane
 plasma soliton \cite{zakharov:1975}.

After substitution $\Psi(x,t)=\psi(x,t) \, e^{i\, t}$ to Eq.\ \eqref{p1kappa} 
we obtain the autonomous equation
\begin{equation}
\label{p1kappa-au}
i \psi_t+\psi_{xx}-\psi+2 |\psi|^{2\,\kappa} \psi = r \psi^{\star}-i \rho \psi,
\end{equation}
which for $\kappa=1$ has been used to describe vertically oscillating layers of water \cite{miles:1984,laedke:1991}, the amplitude modulation of 
nonlinear lattices \cite{denardo:1992}, and the effect of a parametric amplifier in a nonlinear dispersive cavity \cite{longhi:1996}. Furthermore, the envelope of small-amplitude \textit{breathers} in the 
parametrically driven and damped nonlinear 
Klein-Gordon equation (or a Klein-Gordon chain) satisfies Eq.\ \eqref{p1kappa-au}. 
This relationship between the two systems is well established for the case of cubic nonlinearity
\cite{bondila:1995,alexeeva:2000,clerc:2009,valcarcel:2013} using the multiple-scale method.   
It can be extended for higher-order nonlinear Klein-Gordon type equations \cite{saxena:2019,kirrmann:1992} by employing the same approach \cite{dauxois:2010}. 
 Indeed, considering the damped nonlinear Klein-Gordon equation 
\begin{align}
\label{eq:kg}
\phi_{TT}-\phi_{XX}+\left(\frac{\phi}{4} -2 (\kappa+1)  \beta(\kappa) \phi^{2 \kappa+1} \right) [1+8 r \epsilon^2 \cos(2 \Omega T)] &= -\rho \phi_{T},
\end{align}
which is modulated by a periodic force with driving frequency $\Omega=\sqrt{1/4 -\epsilon^2}$ (see Ref. \cite{bondila:1995}), where $\beta(\kappa)>0$, $\epsilon \ll 1$, and $\kappa>0$. The parameters $\kappa$ and $\rho$ play the same roles as in Eq. \eqref{p1kappa-au}. 
It can be shown that the small-amplitude breather solution is given by
\begin{align}
\label{eq:smallb}
\phi(X,T)= 2 \epsilon^{1/\kappa}  \text{Re}(\psi(\epsilon X,\epsilon^2 T) e^{-i \Omega T}),
\end{align} 
where $\beta(\kappa)$ can be chosen such that $\psi(x,t)$ satisfies Eq.\ \eqref{p1kappa-au}. Consequently, it is natural to expect that the results obtained from investigating Eq.\ \eqref{p1kappa-au} can be applied to analyze  the stability of the breather solution \eqref{eq:smallb}.

When $\kappa=1$, Eq.\ \eqref{p1kappa-au} has two exact stationary solutions \cite{barashenkov:1991}. 
Barashenkov, Bogdan and Korobov showed that one of these two solutions is unstable, while the other solution remains stable in a certain region of the plane $\rho$-$r$ \cite{barashenkov:1991}. 
The solution initially stable may become unstable due to a collision of two internal modes (mechanism of oscillatory instability  \cite{alexeeva:1999}).

The aim of the current study is twofold: First, to extend the search for stationary solutions  of Eq.\ \eqref{p1kappa}  
with arbitrary  nonlinearity parameter $\kappa>0$. Second to conduct a stability analysis that explores the dependence on $\kappa$. 
The case $\kappa \ge 2$, where the NLS soliton is unstable in the absence of 
parametric forcing and dissipation, is of particular interest. 
This soliton was stabilized by incorporating an external potential with supersymmetry and parity-time symmetry \cite{cooper:2017}. 
Here, we show that it can  also be stabilized through the combined effect of damping and  parametric force. The unstable soliton may become stable due to the resonance between two internal modes moving in the real axis (mechanism of oscillatory stability).   
 
We found that Eq.\ \eqref{p1kappa} possesses two exact stationary solutions, denoted by $\Psi_{\pm}(x,t)$. Notably, the conditions for existence of $\Psi_{\pm}(x,t)$ are independent of $\kappa$. 
While both solutions share the same functional form, $\Psi_{-}(x,t)$ has a lower amplitude than $\Psi_{+}(x,t)$. We demonstrate that the former solution is always unstable owing to the presence of a positive real eigenvalue in the spectrum of the corresponding Sturm-Liouville problem, which is obtained 
from the linearization of Eq.\ \eqref{p1kappa} around $\Psi_{-}(x,t)$. Subsequently, we provide a parametrization that reduces the parameter space, enabling us to obtain the stability curve, $r(\rho)$, that separates the stable and unstable regions for each value of $\kappa$, including $\kappa \ge 2$. 
 
The paper is organized as follows. In Sec.\ \ref{sec2},  we derive the conditions for the existence of two stationary solutions of Eq.\ \eqref{p1kappa} from the continuity equations and explicitly determine these solutions. Section \ref{sec3} focuses on the linear stability analysis, where we linearize the considered NLS equation around its stationary solutions and obtain the corresponding Sturm–Liouville problem. In Sec.\ \ref{sec4}, we numerically solve the eigenvalue problem and compute the stability curve $r(\rho)$, which delineates the boundary between the stable and unstable regions. In Appendix \ref{appA}, we prove that $\Psi_{-}(x,t)$ is unstable, while in Appendix \ref{appB}, we analytically demonstrate that $\Psi_{+}(x,t)$ exhibits oscillatory instability for $\kappa < 2$ whenever a complex quadruplet emerges. Finally, we summarize the main results and conclusions in Sec.\ \ref{sec5}.

\section{Parametrically driven and damped NLS soliton for $\kappa>0$} \label{sec2}

The nonlinear parameter $\kappa$ and the parametric force explicitly modify the energy density as follows
\begin{align}
\label{eq:energy}
E &=  \int_{-\infty}^{+\infty}  \left(|\psi_x|^2+|\psi|^2-\frac{2}{\kappa+1} |\psi|^{2(\kappa+1)}+ r \frac{ (\psi^\star)^2 + \psi^2}{2} \right) \, dx. 
\end{align}
By inserting the ansatz 
\begin{align}
\label{eq:ssolution}
\psi(x,t)& =\psi_0(x) e^{-\frac{i}{2}
 \,\theta(x)} 
\end{align}
into Eq. \eqref{eq:energy}, it can be verified that the energy, as well as the  momentum 
\begin{align}
\label{eq:momentum}
P = \frac{i}{2} \int_{-\infty}^{+\infty}  (\psi\,\psi_{x}^\star - \psi^\star \, \psi_{x}) \, dx,
\end{align}
and the norm
\begin{align}
\label{eq:norm}
N =  \int_{-\infty}^{+\infty} |\psi|^2 \, dx
\end{align}
are time independent. Moreover, substituting \eqref{eq:ssolution} into the r.h.s of the time variations 
\begin{align}
\label{eq:cenergy}
\frac{dE}{dt}& = \int_{-\infty}^{+\infty} \left\{ 
-i \rho (\psi_t \psi^\star - \psi_{t}^\star \psi)+
i\,r [ (\psi^\star)^2- \psi^2]  \right\} \, dx, \\
\label{eq:cmomentum}
\frac{dP}{dt}& = \int_{-\infty}^{+\infty} \left\{\rho\,i 
(\psi^\star \, \psi_{x}-\psi\,\psi_{x}^\star)+ r\,(\psi^\star\,\psi_{x}^\star + \psi \, \psi_{x})\right\} \, dx,  \\
\label{eq:cnorm}
\frac{dN}{dt}& =\int_{-\infty}^{+\infty}  \left\{ -2 \rho |\psi|^2 +i\,
r [\psi^2- (\psi^\star)^2]\right\} \, dx,
\end{align}
it can be shown that all these integrals vanish 
whenever $\theta(x)=\Theta$ satisfies 
\begin{align}
\sin(\Theta) &=\frac{\rho}{r}.
\end{align}
Therefore, either 
\begin{align}
\label{p5kappa}
\Theta=\Theta_{+}&= \arcsin\left(\frac{\rho}{r}\right), 
\end{align}
or 
\begin{align}
\label{p6kappa}
\Theta=\Theta_{-}&=\pi- \arcsin\left(\frac{\rho}{r}\right).   
\end{align}
In both cases, $\psi_0(x) \equiv \psi(x)$ (hereafter we omit the subscript $0$ for simplicity) satisfies the equation 
\begin{align}
\label{p7kappa}
\psi_{xx}-[1+r \cos(\Theta)] \psi+2 \psi^{2\,\kappa+1} = 0, 
\end{align}
and its solution  is given by   
\begin{align}
\label{p9kappa}
\psi_{\pm}(x)=B_{\pm}  \sech^{1/\kappa}[\kappa\,\sqrt{\omega_{\pm}}\,x], \quad B_{\pm}=\left(\frac{\omega_{\pm} (\kappa+1)}{2}\right)^{1/2\,\kappa},
\end{align}
$\omega_{\pm}\equiv 1+r \cos(\Theta)=1 \pm \sqrt{r^2-\rho^2}>0$, $r \geq\rho$.
Since $\omega_{-}>0$, $\rho  < r<\sqrt{1+\rho^2}$ for $\Psi_{-}(x,t)$. 
Therefore, the two stationary solutions of the 
parametrically driven, damped NLS Eq.\ \eqref{p1kappa} are
\begin{equation}
\label{p10kappa}
\Psi_{\pm}(x,t)=B_{\pm} \sech^{1/\kappa}[\kappa \sqrt{\omega_{\pm}} x] e^{i\,  t-i \Theta_{\pm}/2}.
\end{equation}
For the specific value of $\kappa=1$, this solution agrees with Eqs. (4)-(6) of Ref. \cite{bondila:1995}. Interestingly,  the parameter $\kappa$ influences both the amplitude, and the width of the soliton. Specifically, 
as $\kappa$  decreases, the soliton becomes wider and smaller. However, the two conditions for the  existence of 
these solutions; namely, that their frequency is locked to half the frequency of the parametric drive, and that their phase satisfies Eqs. \eqref{p5kappa}--\eqref{p6kappa}, are independent of $\kappa$.  

A complete stability analysis of this solution requires the variation of $\kappa$, $r$ and $\rho$. 
In the next section, we will provide a linear stability analysis using an exceptional change of variables \cite{barashenkov:1991}, which simplifies the derivation of the curve $r(\rho)$  that separates stable and unstable regions for each value of $\kappa$. 

\section{Linear stability analysis} \label{sec3}

We linearize Eq. \eqref{p1kappa} around its stationary solution  \eqref{p10kappa}, by substituting  $\Psi(x,t)=[\psi_{\pm}(x)+u(x,t)] e^{i \, t-i \Theta_{\pm}/2}$ into Eq.\  \eqref{p1kappa}. Here,  
$\psi_{\pm}(x)$ is given by Eq.\ \eqref{p9kappa}, and the small perturbation $u(x,t)$ can be a  complex function satisfying 
\begin{equation}
\label{v1kappa}
i u_t- u+u_{xx}+2\,(\kappa+1) \psi_{\pm}^{2\,\kappa} u+ 2\,\kappa \psi_{\pm}^{2\,\kappa} u^\star= r e^{i \Theta_{\pm}} u^{\star}-i \rho u.
\end{equation}
Therefore, the real and the complex parts of $u(x,t)=e^{-\rho\,t}\,[f(x,t)+i\,g(x,t)]$ satisfy
\begin{align}
\label{v2kappa}
f_t -\rho f&= \left( -\frac{d^2}{dx^2} - 2 \psi_{\pm}^{2\,\kappa} +1-r\,\cos(\Theta_{\pm}) \right)\,g. \\
\label{v3kappa}
-g_t-\rho\,g &= \left(-\frac{d^2}{dx^2}  -2 (2\,\kappa+1) \psi_{\pm}^{2\,\kappa} +1+r\,\cos(\Theta_{\pm}) \right)\,f. 
\end{align}
With this definition of $u(x,t)$ both equations are influenced by the damping in the same way.
These equations can be rewritten as follows:
\begin{align}
\label{v4kappa}
f_T -\widetilde{\rho} f&= L_0 \,g, \\
\label{v5kappa}
-g_T-\widetilde{\rho}\,g &= L_1\,f,
\end{align}
where the operators $L_0, L_1$ are defined as
\begin{align}
\label{L0}
L_0 &= -\frac{d^2}{dX^2} +1-\varepsilon_{\pm}- (2-\varepsilon_{\pm}) \psi_{\pm}^{2\,\kappa} , \\
\label{L1}
L_1 &=-\frac{d^2}{dX^2}+1  -(2-\varepsilon_{\pm}) (2\,\kappa+1) \psi_{\pm}^{2\,\kappa}, 
\end{align}
and $X=  \sqrt{\omega_{\pm}} x$,  $T=\omega_{\pm} t$, $\varepsilon_{\pm}=2(1-1/\omega_{\pm})$, and $\widetilde{\rho}=\rho/\omega_{\pm}$.  The term 
$(2-\varepsilon_{\pm})\psi_{\pm}^{2\,\kappa}=(\kappa+1)\,\sech^2(\kappa X)$. For $\kappa=1$ these operators agree 
with the studied ones in Ref.\ \cite{barashenkov:1991}.  

Notice that the function $\psi_{\pm}(X)$ satisfies 
\begin{equation}
\label{v6kappa}
\psi_{XX}- \psi+(2-\varepsilon_{\pm}) \psi^{2\,\kappa+1} = 0, 
\end{equation}
with the parameters  $\varepsilon_{-} \in (-\infty,0)$ and $\varepsilon_{+} \in [0,2)$, respectively. 
When  $\varepsilon_{+}=0$, then $r=\rho$, and the solution of this equation reduces to the soliton of the nonlinear Schr\"odinger equation with specific phase.  

 The $\Psi_{-}(x,t)$ solution is found to be always unstable, as is proven in Appendix \ref{appA}. In contrast, to analyze the stability of $\Psi_{+}(x,t)$, we seek the solution of Eqs.\ \eqref{v4kappa}--\eqref{v5kappa} using the following ansatz \cite{barashenkov:1991}.
\begin{align}
\label{eq:8kappa}
f(X,T)&= \exp(\nu T) \Re[\exp(i\Omega T) f_c(X)], \\ 
\label{eq:8akappa}
g(X,T) &= \exp(\nu T) \Re[\exp(i\Omega T) g_c(X)], 
\end{align}
where $\nu$ and $\Omega$ are real numbers and 
$f_c(X)=f_r(X)+i\,f_i(X)$ and $g_c(X)=g_r(X)+i\, g_i(X)$ 
are complex functions that are not identically zero. 
As a consequence of the above ansatz, the soliton will be stable whenever $\nu\leq\widetilde{\rho}$. By inserting these expressions into Eqs.\ \eqref{v4kappa}--\eqref{v5kappa}, and denoting 
$\lambda=\nu+i\Omega$,  we obtain 
\begin{align}
\label{eq:9kappa}
L_0 g_c(X) &= (\lambda-\widetilde{\rho}) f_c(X),  \\
\label{eq:9kappaa}
L_{1} f_c(X) &= - (\lambda+\widetilde{\rho}) g_c(X).
\end{align}
When $\lambda=\nu=\widetilde{\rho}$, the soliton is marginally stable. 
For the case where $\lambda\neq \widetilde{\rho}$, by defining 
\begin{align}
\Lambda^2   &=\lambda^2-\widetilde{\rho}^2, \label{defLambda}\\
\widetilde{g}_c &= \sqrt{\frac{\lambda+\widetilde{\rho}}{\lambda-\widetilde{\rho}}} \, g_c,
\end{align}
the system of Eqs.\ \eqref{eq:9kappa}--\eqref{eq:9kappaa} becomes a 
dissipationless eigenvalue problem
\begin{align}
\label{eq:10kappa}
L_0 \widetilde{g}_c &= \Lambda\, f_c,  \\
\label{eq:10kappaa}
L_{1} f_c & = -\Lambda \widetilde{g}_c.
\end{align}
where $\Lambda$ can be complex.
It has several advantages related to: 
\begin{enumerate}
\item \textbf{Symmetry properties:} The system described by Eqs.\ \eqref{eq:10kappa}--\eqref{eq:10kappaa} is invariant under the transformations where   $\{\Lambda,f_c,\widetilde{g}_c\}$ are replaced by $\{\Lambda^\star,f_c^\star,\widetilde{g}_c^\star\}$. Therefore, if $\Lambda$ is an eigenvalue, $\Lambda^\star$ is also an eigenvalue. Additionally, the  system is invariant under the transformations where $\{\Lambda,f_c,\widetilde{g}_c\}$ are replaced by $\{-\Lambda,f_c,-\widetilde{g}_c\}$. Therefore, if $\Lambda$ is an eigenvalue, $-\Lambda$ is also an eigenvalue.  This implies that the complex eigenvalues $\{-\Lambda,\Lambda,-\Lambda^\star, \Lambda^\star\}$, 
$\Lambda=\Lambda_r+i\Lambda_i$, with $\Lambda_i\Lambda_{r}\neq0$, 
appear in quartets. The value of $\varepsilon$ for which a quadruplet of eigenvalues emerges is  hereafter defined as $\tilde{\varepsilon}$.

\item \textbf{Continuum spectrum:} As $x \to \pm \infty$, Eqs.\ \eqref{eq:10kappa}--\eqref{eq:10kappaa} reduce to a system of two linear equations, with a non-trivial solution given by $f_c \propto \widetilde{g}_c \propto e^{i\,k\,x}$, $\forall k \in \mathbb{R}$, if and only if 
$\Lambda=\Lambda_c(k)= \pm i\,\sqrt{(1+k^2)\,(1+k^2-\varepsilon)}$,
 where, from now on, $\varepsilon=\varepsilon_{+}$. The continuum spectrum lies 
in the intervals $(-\infty,-i\,\sqrt{1-\varepsilon}] \cup [i\,\sqrt{1-\varepsilon},+\infty)$ independent of $\kappa$. 
Clearly, $\Psi_{+}(x,t)$, which exists for $\varepsilon < 2$, could be stable only if $0\le \varepsilon \le 1$, i.e., $\rho \leq r \le \sqrt{1+\rho^2}$). 
Otherwise, the soliton is unstable due to the continuum spectrum.

\item \textbf{Stability Criterion:} Setting $\Lambda = \Lambda_r+i\Lambda_i$
in Eq. \eqref{defLambda}  it follows that
\begin{equation}
\label{eq:11kappa2}
\nu^2 = \frac{\Lambda_r^2-\Lambda_i^2+\widetilde{\rho}^2}{2}+
\frac{\sqrt{(\Lambda_r^2-\Lambda_i^2+\widetilde{\rho}^2)^2+4\,\Lambda_r^2\,\Lambda_i^2}}{2}.
\end{equation}

From \eqref{eq:11kappa2}, we see that for the soliton to be stable $\nu\leq \widetilde{\rho}$, and therefore $\widetilde{\rho}$ and $\Lambda$ should satisfy the following condition 
\begin{equation}\label{stability_condition}
\widetilde{\rho}^2 \left(\Lambda_i^2 - \Lambda_r^2\right) \geq \Lambda_r^2 \Lambda_i^2.
\end{equation}
From this equation  we deduce that a necessary  condition for stability is $|\Lambda_i| > |\Lambda_r|$. In fact, if there exists a $\Lambda$ of the form $\Lambda = i\Lambda_i$, it will never produce instability in the system. However, if $\Lambda = \Lambda_r \neq 0$, then the soliton will be unstable.

\item \textbf{The stability curve:} Assume that for a given $\varepsilon$, 
$\tilde{\varepsilon} \le \tilde{\varepsilon}_c \le \varepsilon \le 1$, a complex eigenvalue 
$\Lambda=\Lambda_r+i\,\Lambda_i$ appears, and moreover, $\Lambda_i> \Lambda_r>0$.  Here, we denote $\tilde{\varepsilon}_c$ the critical value of $\varepsilon$ for which a complex eigenvalue with $|\Lambda_i|>|\Lambda_r|$ arises.
The threshold of instability against a local mode occurs when $\nu=\widetilde{\rho}$, which implies, 
\begin{align} 
\rho(\varepsilon) &= \frac{2\,\nu}{2-\varepsilon}.
\label{eq:11akappa}
\end{align}
where the relation $\widetilde{\rho}=\rho/\omega_{\pm}$ has been used.

By inserting $\Lambda=\Lambda_r+i\,\Lambda_i$, $\lambda=\nu+i\Omega$, and $\nu=\widetilde{\rho}$ into the definition \eqref{defLambda},  we obtain 
\begin{equation} \label{eq:nu}
\nu=\frac{\Lambda_i\, \Lambda_r}{\sqrt{\Lambda_i^2-\Lambda_r^2}},
\end{equation}
where   $\Lambda_i>\Lambda_r$. The values of $\Lambda_r$ and $\Lambda_i$ depend on $\varepsilon$ and $\kappa$.
From the definition of $\varepsilon$, 
the amplitude of the parametric force can be expressed as follows:
\begin{equation}
\label{v7kappa}
r(\varepsilon)=\sqrt{\rho^2(\varepsilon)+\left(\frac{\varepsilon}{2-\varepsilon}\right)^2}.
\end{equation}

  Finally, from Eqs.\ \eqref{eq:11akappa} and \eqref{v7kappa}, we obtain
\begin{align}
\label{eq12kappa}
r(\varepsilon) &= \frac{\sqrt{\varepsilon^2+4\,\nu^2}}{2-\varepsilon}. 
\end{align}
The stability curve is formed by all the points $(\rho(\varepsilon),r(\varepsilon))$, where $\tilde{\varepsilon}_c \le \varepsilon \le 1$. Notice that this analysis holds  for arbitrary  $\kappa$. 
\end{enumerate}

\section{Numerical simulations}\label{sec4}

In order to investigate the stability, first we numerically solve the discrete version of Eq.\ \eqref{v6kappa} by using the Newton-Raphson  
method up to the second or third iteration \cite{press:1992,carretero:2024}. Subsequently, we employ the so-called {\it simplified Newton method}, in which the Jacobian is fixed \cite{ortega:1970}, also referred to as the fixed Newton method \cite{waziri:2010}. Our observations indicate that this method offers  several advantages: 
i) 
it allows for increased accuracy in the numerical solution, achieving a tolerance of $10^{-12}$, 
ii) the Jacobian is computed only during the initial iterations, thus avoiding  the division by a small number that arises when the Jacobian becomes singular and the Newton-Raphson method cannot longer be applied, iii) the numerical solution is consistently centered at zero, and iv) the method converges after few iterations when the exact stationary solution is used as an initial guess.  
 
To discretize the second spatial derivative,  a  second-order central difference scheme is employed. Here, $X \in [-L,L]$, meaning that the nodes are located at $X_j=-L+j\,\Delta X$, $j=0, \cdots, N-1$, and $\Delta X=2\,L/(N-1)$. The numerical simulations were performed using $L=30, 40, 50$ and $N=501, 801, 1001$. Each of these choices ensures that the soliton width is much smaller than the length of the system, $2\,L$, allowing the solitary wave to effectively mimic the stationary solution of the infinite domain. 
Following this, we compute the solution of the eigenvalue problem defined by Eqs.\ \eqref{eq:10kappa}--\eqref{eq:10kappaa} using the same discretization. 
In our numerical code 
\cite{matlab:2020,carretero:2024}, we set $\varepsilon \in [0,1]$ and $0.25 \le \kappa \le 3$. 
To proceed, we will analyze the cases where $\kappa<2$ and $\kappa \ge 2$ separately.

\subsection{$\kappa < 2$}

\begin{figure}[ht!]
\centering\includegraphics[width=0.75\linewidth]{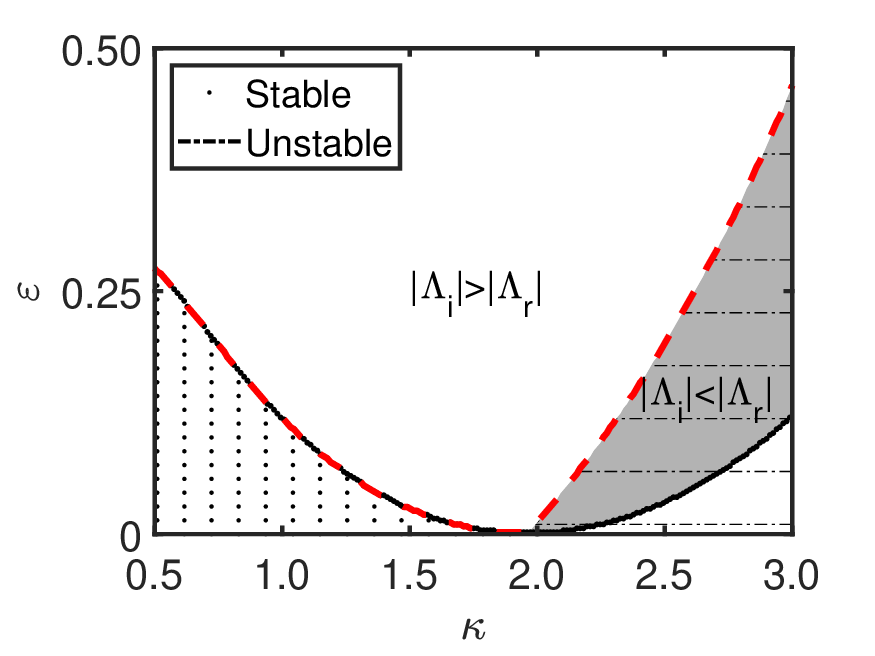} 
\caption{ Critical value of $\tilde{\varepsilon}_c$ as a function of $\kappa$ is depicted with a red dashed line.   In the region above this curve, the necessary condition for stability $|\Lambda_i|>|\Lambda_r|$ holds. Black solid line (overimposing with red dashed line for $\kappa<2$) represents the values of 
$\tilde{\varepsilon}$ versus $\kappa$ for which a quadruplet emerges. 
The dotted region is stable since $\Lambda_r=0$. In the shadow region the condition 
$|\Lambda_i|<|\Lambda_r|$ is satisfied. In the dot-dashed region the necessary condition for stability  \eqref{stability_condition} is not satisfied and 
the solution is unstable. 
Parameters: $L=50$ and $N=501$.}
\label{fig1}
\end{figure}

It is observed that for fixed  $\kappa<2$ and small $\varepsilon$, all eigenvalues are either close to  zero (of order of $10^{-7}$, $10^{-8}$) or purely  imaginary, implying the stability of the soliton  (see the dotted region in Fig.~\ref{fig1}). Fixing $\kappa$, there exists a value of $\varepsilon=\tilde{\varepsilon}$ for which a quadruplet of eigenvalues $\{\Lambda,-\Lambda,\Lambda^\star,-\Lambda^\star\}$ ($\Lambda=\Lambda_r+i\Lambda_i$) emerges (black solid line in Fig.~\ref{fig1}). 
In the appendix B for $\varepsilon > \tilde{\varepsilon}$ we analytically prove that
$|\Lambda_i| > |\Lambda_r|$ for $\kappa < 2$, and, therefore, $\tilde{\varepsilon} = \tilde{\varepsilon}_c$
 This is numerically confirmed in Fig.~ \ref{fig1}, where 
 the red dashed line  ($\tilde{\varepsilon}_c(\kappa)$) is overimposed with the black solid line ($\tilde{\varepsilon}(\kappa)$).

\begin{figure}[ht!]
\centering
\begin{tabular}{cc}
\includegraphics[width=0.5\linewidth]{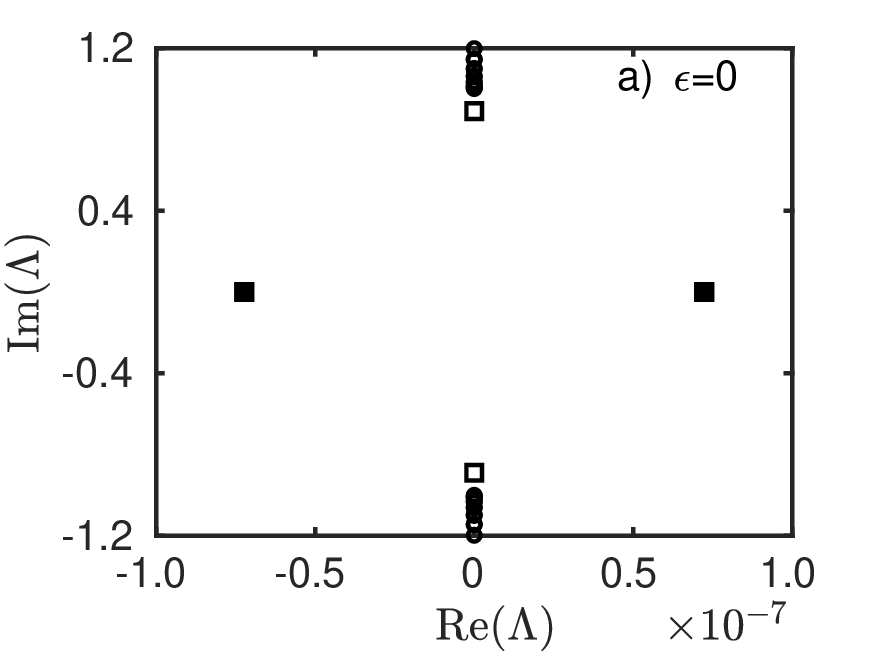} & \includegraphics[width=0.5\linewidth]{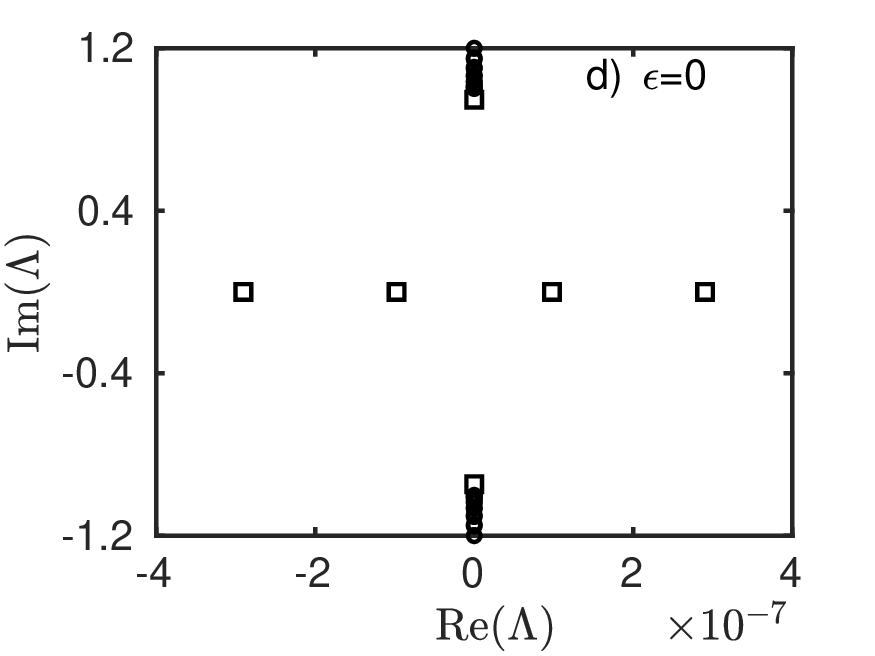} \\
\includegraphics[width=0.5\linewidth]{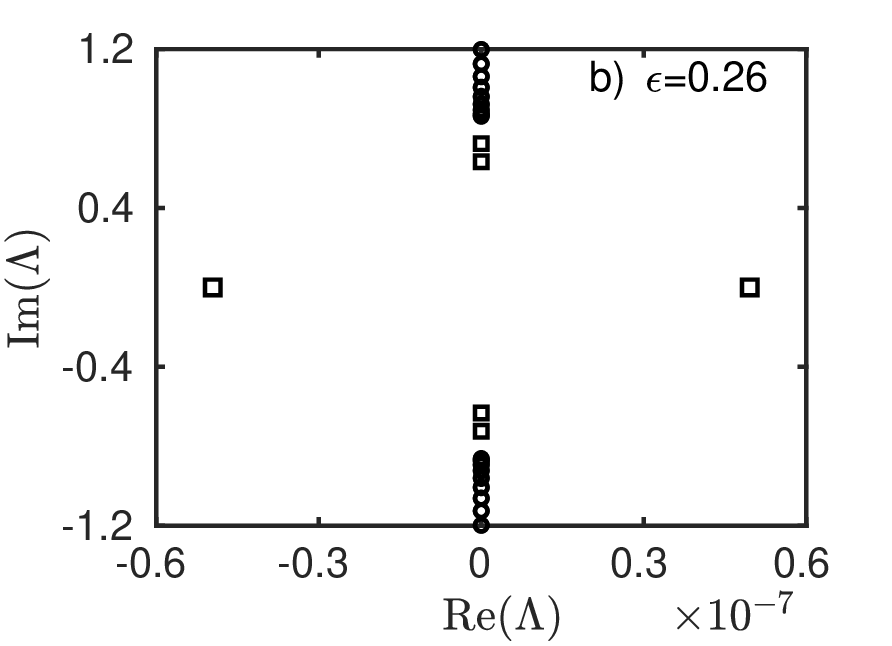} & \includegraphics[width=0.5\linewidth]{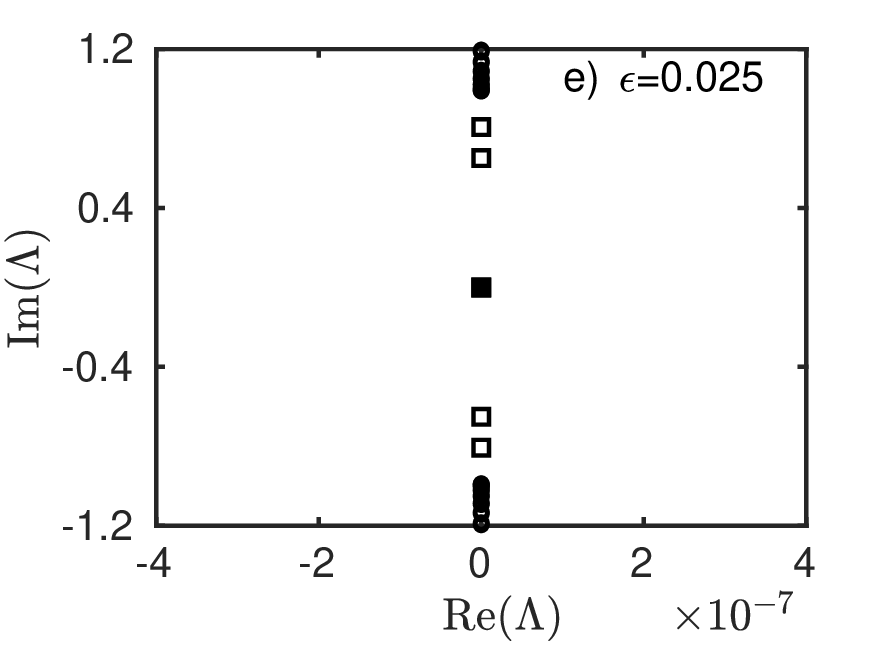}\\
\includegraphics[width=0.5\linewidth]{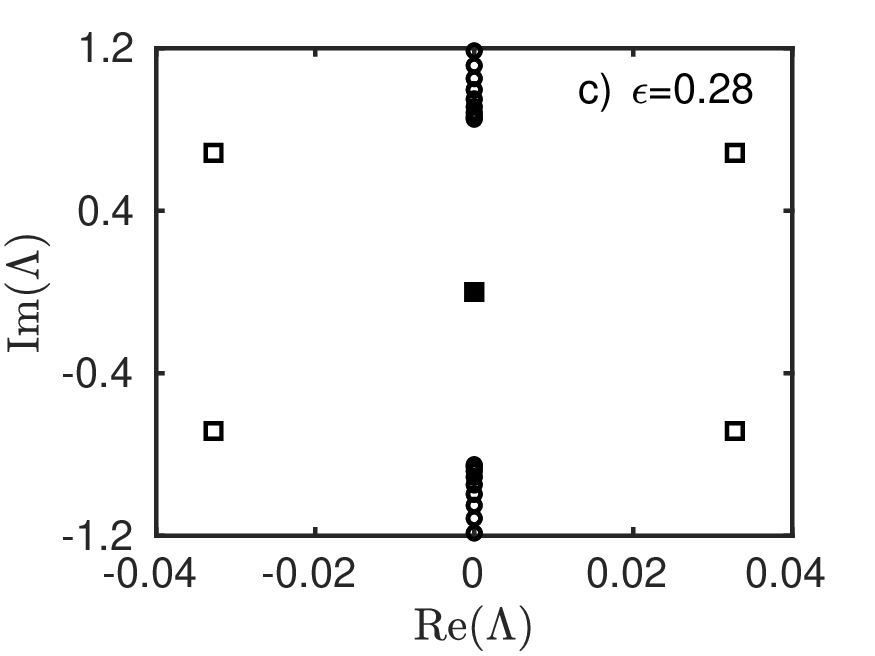} & \includegraphics[width=0.5\linewidth]{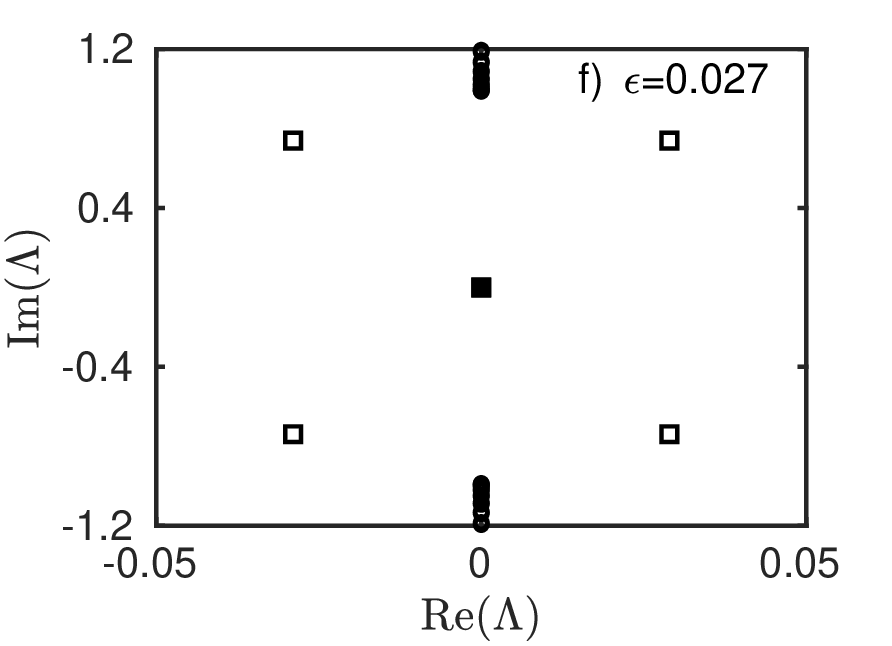} 
\end{tabular}
\caption{Crucial eigenvalues illustrating the 
emergence of complex quadruplet as $\varepsilon$ increases. Left-hand side panel: $\kappa=0.5$ ($\tilde{\varepsilon}_c=0.273$).  
Right-hand side panel: 
$\kappa=1.5$ ($\tilde{\varepsilon}_c=0.0268$).  
 Circles: several eigenvalues of the continuous spectrum. Squares: discrete eigenvalues. 
The filled squares represent two very close eigenvalues near zero.
Parameters: $L=50$ and $N=1001$.
}
\label{fig2}
\end{figure}

In Fig.~\ref{fig2}, the crucial eigenvalues are represented for two specific cases: $\kappa=0.5<1$ and $\kappa=1.5>1$. For $\varepsilon =0$, Fig.~\ref{fig2}a) and \ref{fig2}d),  
the discrete spectrum contains a mode situated in the gap between the zero modes and the continuum spectrum. 
In the middle panels, ~\ref{fig2}b) and \ref{fig2}e), as $\varepsilon$ is increased ($\varepsilon<\tilde{\varepsilon}_c$), the absolute value of the discrete mode decreases, approaching that of a detached mode from zero. These two values can be observed in the  vertical axis of the Fig.~\ref{fig2}. Further increment of $\varepsilon$ beyond $\tilde{\varepsilon}_c$ leads to the appearance of a complex quadruplet [visible in Fig.~\ref{fig2}c) and Fig.~\ref{fig2}f)] following the collision of these two modes. This instability mechanism is analogous to the case of $\kappa=1$ \cite{barashenkov:1991}, which lacks an internal mode when $\varepsilon=0$  
\cite{scott:2003,dauxois:2010}. 
Let us remark that since the numerical values of zero-eigenvalues are not exactly zero, we 
observe in Fig.\ \ref{fig2}a) four eigenvalues near zero $(\pm 7.23 \pm 8.62\,i) \times 10^{-8}$, which approach zero as $N$ increases (due to the scale only two are visible). In Fig. \ref{fig2}b) 
the two eigenvalues near zero are $\pm 4.96 \times 10^{-8}$, while in  Fig. \ref{fig2}c) 
these two eigenvalues lie in the vertical axis $\pm 1.61\,i \times 10^{-8}$. 
For $\kappa=1.5$ we have a similar situation with the zero-eigenvalues: in Fig. 
\ref{fig2}d) we observe four eigenvalues $\pm 9.8 \times 10^{-8}$ and $\pm 2.9 \times 10^{-8}$, 
and in Fig.\ \ref{fig2}e) and f), the two eigenvalues near to zero are 
$\pm 7.8\,i \times 10^{-8}$ and $\pm 1.01\,i \times 10^{-7}$, respectively.

\begin{figure}[ht!]
\centering\includegraphics[width=0.75\linewidth]{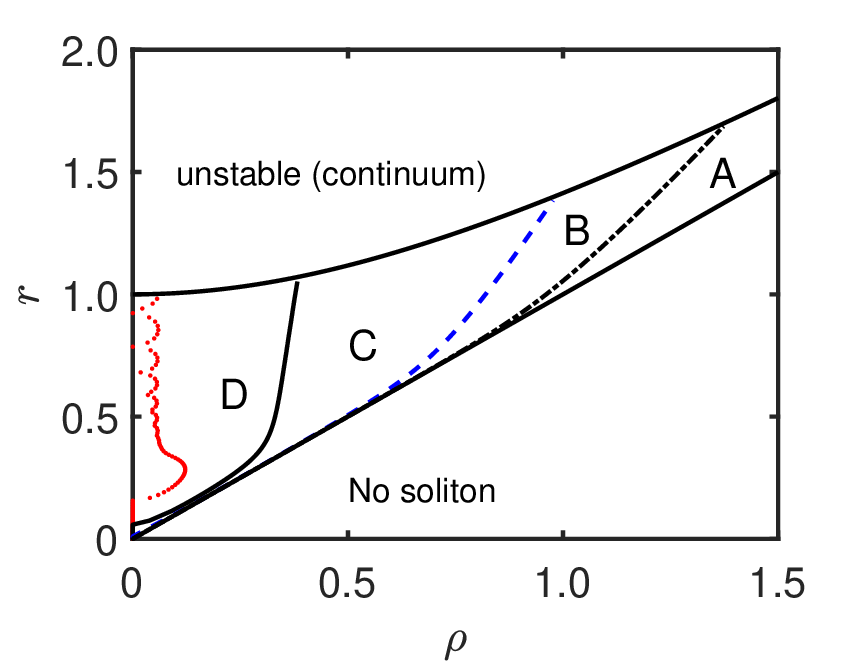} \\
\centering\includegraphics[width=0.75\linewidth]{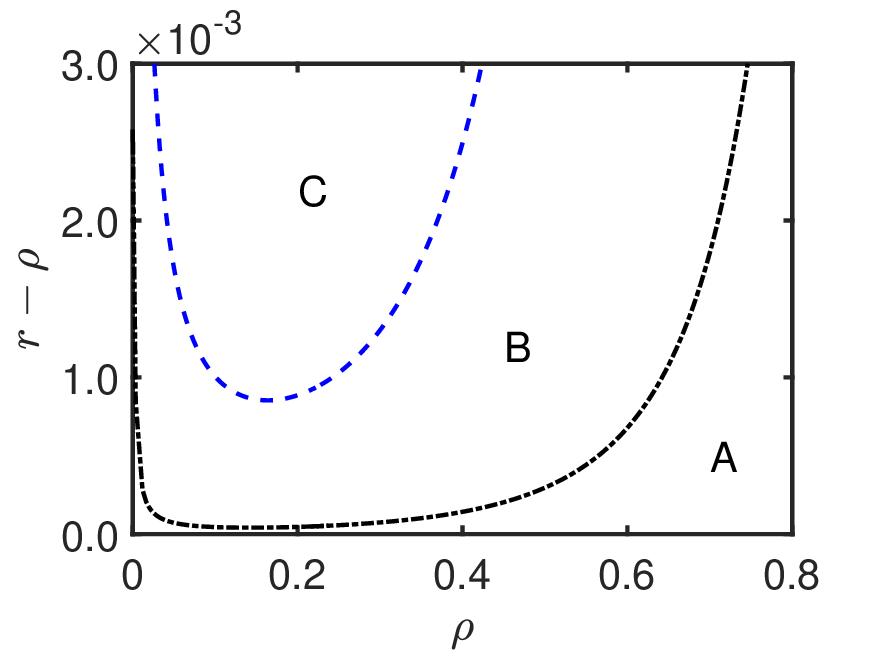}
\caption{Upper panel: The stability curves, traveled from the lower to the upper boundaries as $\varepsilon$ increases ($\tilde{\varepsilon}_c \le \varepsilon \le 1$),  are depicted for several values of $\kappa<2$: a red dotted line  for $\kappa=0.5$, a black solid line for $\kappa=1$, a blue dashed line  for $\kappa=1.5$, and a  dot-dashed black line for  $\kappa=1.75$.
These curves are bounded below by $r=\rho$ and above by $r= \sqrt{1+\rho^2}$, both represented by  black solid lines. 
For each value of $\kappa$, the soliton is stable when the point $(\rho,r)$ lie on the region to the right-hand side of the corresponding curve. Otherwise, it is deemed unstable. 
Lower panel: For $\kappa=1.5$ and $\kappa=1.75$, it is represented $r(\rho)-\rho$ versus $\rho$, where $r(\rho)$ is the stability curve.   
Parameters: $L=50$ and $N=501$.}
\label{fig3}
\end{figure}

Upon the emergence of the quadruplet  with $|\Lambda_i| > |\Lambda_r|$,  $\rho(\varepsilon)$ and $r(\varepsilon)$ are computed  using Eqs. \eqref{eq:11akappa}, \eqref{eq:nu} and \eqref{eq12kappa}, respectively. 
These values correspond to the stability curve, which separates the stable and unstable regions. It is represented in Fig.~\ref{fig3} for various values of $\kappa$. These curves are traveled from the lower limit close to $r=\rho$ ($\varepsilon=0$) to the upper limit $r=\sqrt{1+\rho^2}$ ($\varepsilon=1$).  
 In this way,  the sense of traveling is defined.

For each value of $\kappa$, the soliton is stable for the parameters $(\rho,r)$ that lie in the region to  the right-hand side of the traveled stability curve $r(\rho)$. Otherwise, it is unstable. These curves monotonically increase with the damping coefficient, except for some values of $\kappa<1$. For instance, for the  case of $\kappa=0.5$ shown in Fig.~\ref{fig3} the monotonicity changes. 
The monotonicity of these curves is related to the motion of the quadruplet  as  
$\varepsilon \ge \tilde{\varepsilon}_c$ is increased. In supplemental videos \cite{supm:2024}, 
a comparison of the quadruplet's motion for $\kappa=1.5$ and $\kappa=0.5$ reveals that in the former case,  the eigenvalues move apart as $\varepsilon$ is increased, and we move along the curve. In contrast,  
for $\kappa=0.5$, this behavior is initially observed; however, after a certain value of $\varepsilon$, the 4 eigenvalues begin to attract  each other.  
The parameters used in the simulations depicted in the videos are $L=50$, $N=501$, and the Jacobian is computed only during the first iteration, since it becomes singular very quickly for $\kappa<1$.  
 
Another interesting consequence of the change in monotonicity is the stabilization of the soliton as $r$ increases.
Specifically, for $\kappa=0.5$, and a fixed damping coefficient $\rho=0.09$, we observe that for $r=0.27$ the soliton is unstable; however as the amplitude of the parametric force is increased to $r=0.4$, the soliton recovers its stability. 
 
Stability diagrams in  the upper panel of Fig.~\ref{fig3} show how the stability can be controlled by the nonlinearity parameter $\kappa$. Generally, it can be achieved by decreasing $\kappa$.
Indeed,  moving from region A to B, then to C, and finally to D, 
we find that on region A,  the stationary solution is stable for $\kappa=0.5, 1.0, 1.5, 1.75$. 
On region B, the solution is stable for $\kappa=0.5, 1.0, 1.5$ and unstable for $\kappa=1.75$. On region C, it is stable for $\kappa=0.5, 1.0$ and unstable for $\kappa=1.5, 1.75$. Finally, on region D, stability takes place when $\kappa=0.5$. 
 In the lower panel of Fig.~\ref{fig3} it is represented $r(\rho)-\rho$ as a function of $\rho$ for two values of $\kappa=1.5$ (blue dashed line) and $\kappa=1.75$ (dot-dashed black line). The region below these curves show that for each value of $\kappa$, there is a stable corridor between the dashed curves and the $r=\rho$ line in the upper panel of Fig.~\ref{fig3}. For $\rho=0$ and $\kappa=1.5$, the threshold value of $r$ from which  
the soliton becomes unstable is $0.013$. This value is reduced to $0.0026$ for $\kappa=1.75$.
    
\subsection{$\kappa \ge 2$}

 For $\varepsilon=0$ we recover the NLS equation with nonlinearity parameter $\kappa$. It is well known 
that for $\kappa \ge 2$, the stationary NLS solution suffers a blowup of the wave amplitude  and becomes unstable \cite{sulem:1999}. Here, we numerically investigate whether the additional parametric force and damping can stabilize $\Psi_{+}(x,t)$. We employ the same numerical schemes,  and solve the discrete versions of Eq.\ \eqref{v6kappa} and eigenvalue problem \eqref{eq:10kappa}--\eqref{eq:10kappaa}, but now setting $2 \le \kappa \le 3$. 

\begin{figure}[ht!]
\centering\includegraphics[width=0.75\linewidth]{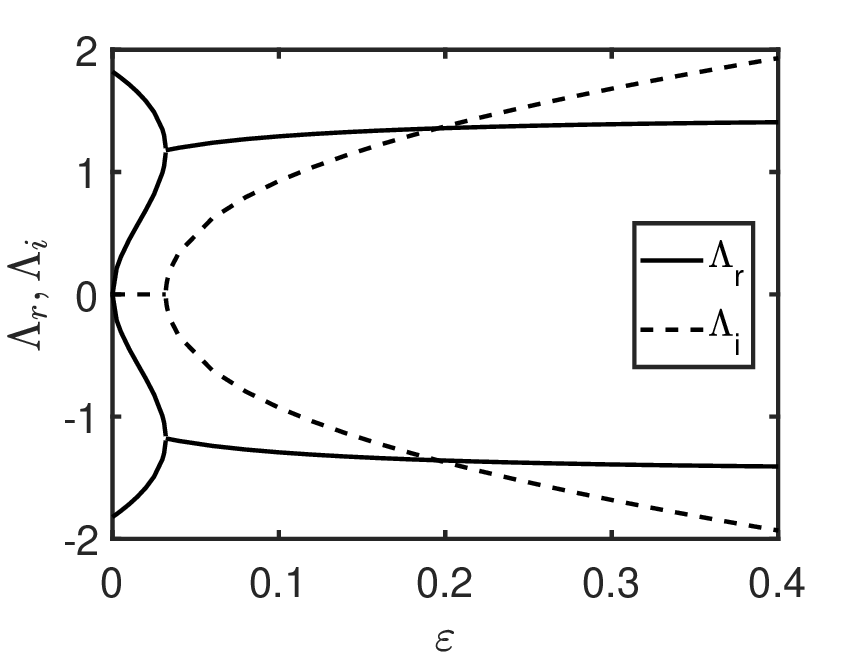} 
\caption{For $\kappa=2.5$, the real and imaginary parts of the eigenvalues which determine the soliton stability are represented as a function of $\varepsilon$. Parameters: $L=50$ and $N=501$.}
\label{fig4}
\end{figure}

\begin{figure}[ht!]
\centering\includegraphics[width=0.75\linewidth]{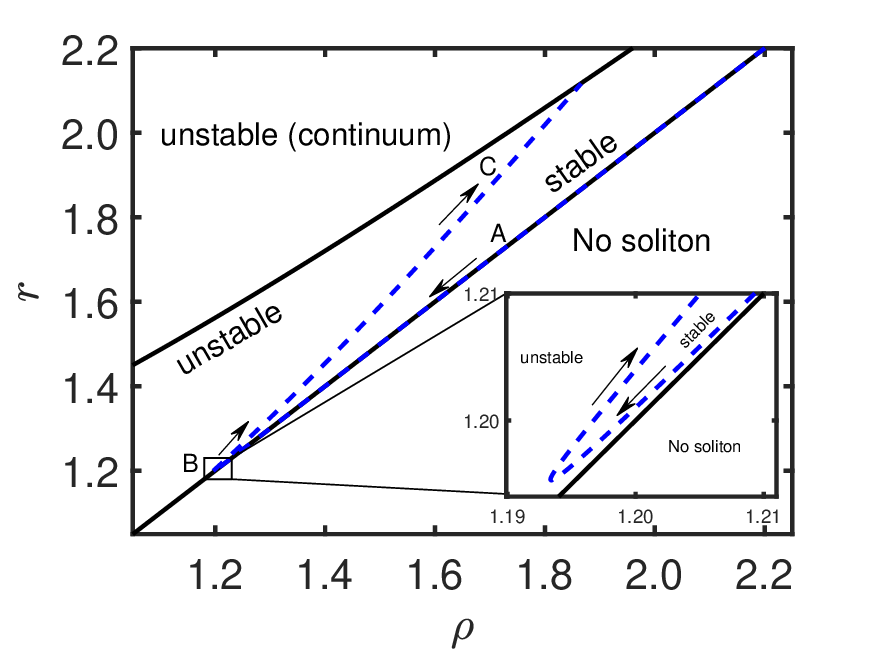} 
\includegraphics[width=0.75\linewidth]{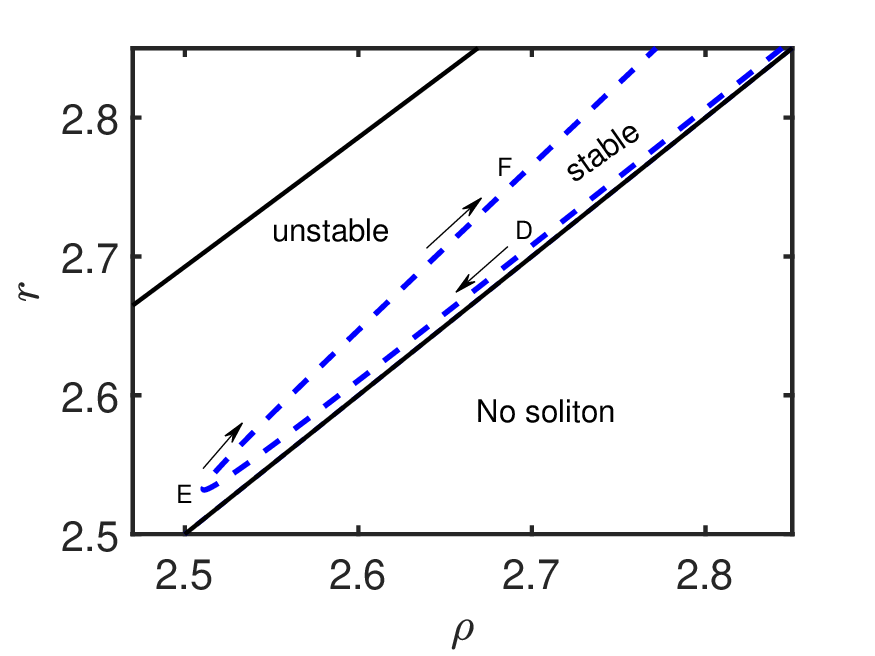} 
\caption{
The stability curves  are depicted with blue dashed line for $\kappa=2$ (upper panel) and $\kappa=2.5$ (lower panel). The curves are bounded below by $r=\rho$ and above by $r= \sqrt{1+\rho^2}$, both represented by  black solid lines. The arrows indicate the direction of movement as 
$\epsilon$ increases: from A to C through B in the upper panel, and from D to F through E in the lower panel. 
For each value of $\kappa$, the soliton is stable when the point $(\rho,r)$ lies in the region to the right-hand side of the corresponding traveled curve. Otherwise, it is deemed unstable. Parameters: $L=50$ and $N=501$.}
\label{fig5}
\end{figure}

Fixing $\kappa$, when $\varepsilon$ is increased, we numerically identify in the spectrum four eigenvalues of order $ O( 10^{-6})$, and other pair of real eigenvalues $\Lambda=\pm \Lambda_r$. 
Therefore, the solitary wave is unstable (see the dot-dashed region below the black solid line in Fig.~\ref{fig1}). 
As $\varepsilon$   increases the real eigenvalues move to zero and two zero  eigenvalues move toward them until they collapse and a complex quadruplet  arises at $\varepsilon=\tilde{\varepsilon}$ 
(see the black solid line in Fig.~\ref{fig1}). In Fig.~\ref{fig4}, for $\kappa=2.5$,  
the formation of a complex quadruplet is represented at $\varepsilon=\tilde{\varepsilon}$, where the dashed line bifurcates in two branches.   In the interval $\tilde{\varepsilon} \le \varepsilon <\tilde{\varepsilon}_c$ the solution remains unstable  (the condition  $|\Lambda_i|<|\Lambda_r|$ holds for $\kappa \ge 2$ and defines the shadow region represented between the black solid line and red dashed line in Fig.~\ref{fig1}). For more details, see also a video  in the supplementary material \cite{supm:2024}. 
  At $\varepsilon=\tilde{\varepsilon}_c\approx 0.2$, $\Lambda_r=\Lambda_i$ (see the intersection between the solid and dashed lines in Fig.~\ref{fig4}). A  further increment of $\varepsilon$ beyond $\tilde{\varepsilon}_c$ leads to the condition   $|\Lambda_{i}| > |\Lambda_r|$  which allows  the stability curve to be computed. In Fig.~\ref{fig5} the stability curves for $\kappa=2$ and $\kappa=2.5$ are shown. 
  Notice that these curves have two branches, and the path moves 
  from the lower branch to the upper branch. For instance, for $\kappa=2$ (see the inset in the upper panel of Fig.\ \ref{fig5}), at $\epsilon=0.019$, we obtain point A of the stability curve at $\rho_A=1.74192$ and $r_A=1.74194$. As $\epsilon$ increases, the path follows the lower branch from A toward B. Point B, located at $\rho_B=1.1933$ and $r_B=1.1953$ is reached at $\epsilon=0.129$. 
A further increase in $\epsilon$ up to $0.893$ enables movement along the upper branch from B toward C, where  $\rho_C=1.7166$ and $r_C=1.8967$. A similar situation is observed in the lower panel of Fig. \ref{fig5} for $\kappa=2.5$, where the curve is traversed from D to F through E. In this case, for  $\epsilon=0.3449, 0.4950$, and $0.7350$ the points D $(2.6932,2.7013)$, E $(2.5106,2.5321)$, and F  $(2.6816,2.7438)$ are obtained, respectively.
   This indicates that the soliton is stable in the region between these two branches. 

\section{Summary} \label{sec5}

We have investigated the parametrically driven, damped nonlinear Schr\"odinger equation with an arbitrary nonlinearity parameter $\kappa>0$, focusing on the control of stability through nonlinearity. 
Two exact stationary solutions, $\Psi_{\pm}(x,t)$, were found. 
These solutions exist when their frequency is locked to half of the parametric force's frequency. They are constrained by specific phase conditions, determined by the damping coefficient $\rho$ and by the amplitude $r$ of the parametric force. These conditions are independent of $\kappa$.

The linear stability of these waves was examined by linearizing the parametrically, driven NLS equation around $\Psi_{\pm}(x,t)$. This approach leads to a separation of variables, resulting in a  Sturm-Liouville problem (SLP) that depends on the parameters $r$, $\rho$ and $\kappa$. 
Consequently, there are three ways to modify the spectrum of this problem and thus its stability: varying  $r$, modifying $\rho$,  and by adjusting the additional parameter $\kappa$. 

The presence of these multiple parameters enriches the analysis while complicating the 
resolution of the problem. A special parametrization through $\varepsilon=\varepsilon_{\pm}$ provides an   efficient and precise method for determining the stability. 
For $\Psi_{-}(x,t)$, where $\varepsilon_{-} \in (-\infty,0)$, we  analytically demonstrate that the spectrum of SLP contains a real positive eigenvalue, indicating that the solution is unstable for all values of $\kappa$. 

Conversely, for $\Psi_{+}(x,t)$, where $\varepsilon_{+} \in [0,1]$, the SLP is numerically solved. Our primary findings are as follows: 
i) for $0.25 \le \kappa<2$, there exists a critical value of $\tilde{\varepsilon}_c$ such that when  
$\varepsilon_{+}<\tilde{\varepsilon}_c$, the spectrum contains  numerically close to zero and pure imaginary eigenvalues, indicating that the soliton is stable. For $\kappa = 1$ the results obtained in Ref.  \cite{barashenkov:1991} are recovered. 
ii) As $\varepsilon_{+}$ increases,  the internal mode and a detached mode from zero move toward each other along the imaginary axis. At the critical value $\varepsilon_{+}=\tilde{\varepsilon}_c$, they  collapse, 
leading to the formation of a quadruplet $\{\Lambda,-\Lambda,\Lambda^\star,-\Lambda^\star\}$, $\Lambda=\Lambda_r + i\Lambda_i$  with a repulsion among their eigenvalues.  
iii) We demonstrate that $|\Lambda_i|>|\Lambda_r|$ when $\kappa<2$. As a consequence,  we  obtain a family of $\kappa$-stability curve, $r(\rho)$, which defines the boundary between the unstable and stable regions. This mechanism resembles the oscillatory instability observed when $\kappa = 1$, where the role of the internal mode is taken over by a mode that has separated from the continuum spectrum \cite{barashenkov:1991}. iv) 
We investigate  several cases in detail, specifically for $\kappa<1$, and $1<\kappa<2$. In the former case, 
the monotonicity of the curves $r(\rho)$ undergoes a change owing to the  attraction of a complex quadruplet toward the continuum spectrum.  

Additionally, we numerically assess the stability of $\Psi_{+}(x,t)$  for  $2 \le \kappa \le 3$. For $\varepsilon_{+}=0$, the soliton is unstable due to the presence of  
 a pair of nonzero real eigenvalues in the spectrum, consistent with previous studies on amplitude blowup in the NLS equation \cite{sulem:1999}. 
As $\varepsilon_{+}$ increases, this pair of eigenvalues moves toward zero, while another pair moves away from zero toward them. The collision of these eigenvalues occurs in the real axis. Although a complex quadruplet emerges, the soliton remains unstable since $|\Lambda_i| \le |\Lambda_r|$. Solely when  $\varepsilon_{+} > \tilde{\varepsilon}_c$ and $|\Lambda_i| > |\Lambda_r|$ a stability region arises. 

In conclusion, the parametric force stabilizes the damped soliton, even for $\kappa \ge 2$ where oscillatory stability is observed; however, it simultaneously breaks Galilean invariance, preventing the derivation of moving solitons from stationary solutions. Future work may focus on analyzing the existence and stability of moving solitons in the parametrically driven, damped nonlinear Schr\"odinger equation with arbitrary nonlinearity parameter.  
Our results can also be of interest in the search for stable \textit{breathers} in the nonlinear Klein-Gordon equations with higher-order nonlinearities.   

\appendix
\section{ $\Psi_{-}(X)$ is unstable} \label{appA}

Here we prove that $\Psi_{-}(X)$ given by Eq.\ \eqref{p9kappa} is 
always unstable solution regardless of the value of $\kappa>0$. 
In order to consider this solution we  assume 
$\varepsilon=\varepsilon_-<0$. For  $\kappa=1$ this was 
proved in Ref.\cite{barashenkov:1990}. We will need the following two propositions:\medskip

\noindent\textbf{Proposition A1:} For any $\varepsilon$ and  $\kappa>0$, 
$L_1$ possesses a zero eigenvalue associated with the eigenfunction
$\psi_X(X)$, which implies that $L_1$ also has always a negative eigenvalue associated with a nodeless eigenfunction, $\psi_0(X)$. 

\noindent\textbf{Proof}: 
From Eqs.\ \eqref{L1} and \eqref{v6kappa} follows that $L_1 \psi_X = 0$. 
This implies that $\psi_X(X)$ is an eigenfunction of $L_1$ associated with the 
zero eigenvalue. Since $L_1$ is a Sturm-Liouville operator, its eigenvalues are ordered in increasing magnitude (see e.g. Ref. \cite[page 722]{morse:1953}), corresponding to the number of nodes in their respective eigenfunctions. 
This implies that if $\psi_X(X)$ has one node, then there exists a nodeless eigenfunction, $\psi_0(X)$, associated with a negative eigenvalue. \hfill $\Box$\medskip

\noindent\textbf{Proposition A2:} $L_0$ is strictly positive defined for $\varepsilon <0$.

\noindent\textbf{Proof}: From  Eq. \eqref{v6kappa} it follows that $L_0 \psi = -\varepsilon \psi$. Because $\psi(X)$ is a nodeless function, their corresponding eigenvalue $-\varepsilon$ 
is the lowest one (see e.g. \cite[page 722]{morse:1953}). Since $\varepsilon <0 $, then the lowest eigenvalue is positive, and then the others eigenvalues are also positive.
Moreover, for the continuous spectrum of $L_0$, $\sigma_c(L_0)\subset[\sqrt{1-\varepsilon},+\infty)$. Therefore, by Theorem 2.18 of \cite{teschl:2009}, for $\varepsilon<0$, $L_0$ is strictly  positive defined operator.  \hfill $\Box$\medskip

Since the operator $L_0$ is positive 
definite (Proposition A2) and therefore invertible, then the system \eqref{eq:10kappa}--\eqref{eq:10kappaa}
can be rewritten as
\begin{equation}\label{12}
L_1 f_c = -\Lambda^2 L_0^{-1} f_c.
\end{equation}
Thus
\begin{equation} \label{eqLambda}
\Lambda^2 = -\frac{\langle f_c | L_1 f_c \rangle}{\langle f_c | L_0^{-1} f_c \rangle}.
\end{equation} 
Let us now define the functional $\Sigma : H^1(\mathbb{C}) \to \mathbb{R}$ as 
\begin{equation}\label{fun-sig}
\Sigma(\xi) = -\frac{\langle \xi | L_1 \xi \rangle}{\langle \xi | L_0^{-1} \xi \rangle},
\end{equation}
where $H^1(\mathbb{C}) := \{\xi \in L^2(\mathbb{C}), \partial \xi \in L^2(\mathbb{C})\}$, where $\partial$ denotes the weak derivative. Since $L_0$ is strictly positive definite, the denominator of the functional $\Sigma$ given in Eq.\ \eqref{fun-sig} does not vanish, and since $L_0$ and $L_1$ are hermitian operators, the functional
is well defined.

Let $\Lambda_0^2$ be  
\begin{equation} \label{34}
\Lambda_0^2 = 
\max_\xi \Sigma(\xi).
\end{equation} 
Taking into account that $\Sigma(\psi_0(X)) >0$, where $\psi_0(X)$ is the fundamental eigenfunction of $L_1$ (Proposition A1), we obtain that $\Lambda_0^2>0$. 

The next step is to verify if $\Lambda_0^2$ is an eigenvalue of Eq.\ \eqref{12}. Let's denote as $\xi_0$ the function where $\Sigma(\xi_0) = \Lambda_0^2$, and choose $\alpha \in \mathbb{R}$ and $h\in H^1(\mathbb{C})$. 

Since the functional $\Sigma$ attains its maximum at $\xi_0$, then we have that
\begin{equation*}
0 = \left.\frac{d}{d\alpha}\Sigma(\xi_0+\alpha h)\right|_{\alpha = 0} =  -\frac{\left(\langle h | L_1 \xi_0 \rangle+\langle \xi_0 | L_1 h \rangle\right) \langle \xi_0 | L_0^{-1} \xi_0 \rangle-\langle \xi_0 | L_1 \xi_0 \rangle \left(  \langle h | L_0^{-1} \xi_0 \rangle+ \langle \xi_0 | L_0^{-1} h \rangle\right)}{\langle \xi_0 | L_0^{-1} \xi_0 \rangle^2},
\end{equation*}
and thus
\begin{align*}
0 = & \left(\langle h | L_1 \xi_0 \rangle+\langle \xi_0 | L_1 h \rangle\right)+\Lambda_0^2 \left(  \langle h | L_0^{-1} \xi_0 \rangle+ \langle \xi_0 | L_0^{-1} h \rangle\right) \\
=  & \langle h | \left(L_1 +\Lambda_0^2 L_0^{-1}\right) \xi_0 \rangle+\langle \xi_0| \left(L_1 +\Lambda_0^2 L_0^{-1}\right) h \rangle\\
=  & \langle h | \left(L_1 +\Lambda_0^2 L_0^{-1}\right) \xi_0 \rangle+\langle h| \left(L_1 +\Lambda_0^2 L_0^{-1}\right) \xi_0 \rangle^*=2\Re \langle h | \left(L_1 +\Lambda_0^2 L_0^{-1}\right) \xi_0 \rangle.
\end{align*}
Because $h$ is any function in the space, we deduce that  
\begin{equation}
\left(L_1 +\Lambda_0^2 L_0^{-1}\right) \xi_0 = 0,
\end{equation}
which proves that $\Lambda_0^2>0$ satisfies Eq.\ \eqref{12}.

This implies that there exists $\Lambda_0 \in \mathbb{R}^+$ in the spectrum 
of the system \eqref{eq:10kappa}--\eqref{eq:10kappaa}, and so we have that for $\varepsilon<0$, the solution  $\psi_{-}(X)$ is always unstable.

\section{$|\Lambda_{i}| > |\Lambda_{r}|$ for $\kappa<2$} \label{appB}

In the resolution of the eigenvalue problem corresponding to the stability analysis of $\Psi_{+}(x,t)$, when $\varepsilon \in [\tilde{\varepsilon},1]$, $\tilde{\varepsilon} >0$, appear four complex eigenvalues $\{\Lambda,-\Lambda,\Lambda^\star,-\Lambda^\star\}$, being $\Lambda=\Lambda_r + i \Lambda_i$. 
Here, we will prove that if $\kappa<2$, then $|\Lambda_{i}| > |\Lambda_{r}|$. This implies that  $\tilde{\varepsilon}=\tilde{\varepsilon}_c$ for $\kappa<2$, where $\tilde{\varepsilon}_c$ is the threshold value for the emergence of quadruplet.  

Let us consider the operator $\widetilde{L}_0 =  L_0+\varepsilon$, where 
$L_0$ is defined by Eq. \eqref{L0} and $\widetilde{L}_0$ satisfies 
\begin{equation}\label{psi}
\widetilde{L}_0\psi =\left(L_1 + 2\kappa(2-\varepsilon)\psi^{2\kappa}\right)\psi = 0.
\end{equation}
Next, we will consider the following functions
\begin{equation}
f_c = \beta \psi + \bar{f}, \qquad \widetilde{g}_c = \alpha \psi + \bar{g},
\end{equation}
where both $\bar{g}$ and $\bar{f}$ are orthogonal to $\psi$, and $\alpha, \beta \in \mathbb{C}$.
Inserting them into the system \eqref{eq:10kappa}--\eqref{eq:10kappaa}, we obtain
\begin{align}
\label{eqn1a}
\widetilde{L}_0 \bar{g} -\varepsilon\bar{g} &= (\varepsilon \alpha+\beta \Lambda)\psi + \Lambda \bar{f}, \\
\label{eqn1b}
{L}_1 \bar{f} + L_1 \beta \psi &= -\Lambda \alpha \psi - \Lambda \bar{g} .
\end{align}
Multiplying  Eq. \eqref{eqn1a} by $\psi$ and integrating over $\mathbb{R}$ we get
\begin{equation}
\begin{split} \label{eqn2}
\langle \psi |\widetilde{L}_0 \bar{g} \rangle-\varepsilon\langle \psi |\bar{g} \rangle= (\varepsilon\alpha+\beta \Lambda)\langle \psi |\psi \rangle+ \Lambda \langle \psi |\bar{f} \rangle, 
\end{split}
\end{equation}
from where, taking into account that $\langle \psi |\widetilde{L}_0 \bar{g} \rangle = \langle \widetilde{L}_0  \psi |\bar{g} \rangle = 0$, Eq.  \eqref{psi}, as well as 
the orthogonality of $\bar{g}$ and $\bar{f}$ with respect to $\psi$, we find the relation $\varepsilon\alpha  = -\beta \Lambda$. Using this relation  Eq. \eqref{eqn1a} becomes 
\begin{align} \label{eqn3a}
\widetilde{L}_0 \bar{g} -\varepsilon\bar{g} &= \Lambda \bar{f}. 
\end{align}
Now we multiply  Eq. \eqref{eqn3a} by $\bar{g}$ and integrate over $\mathbb{R}$ to get
\begin{equation}
\langle \bar{g} |\widetilde{L}_0 \bar{g} \rangle -\varepsilon\braket{\bar{g} }{\bar{g} }= \Lambda \braket{\bar{g} }{\bar{f} }. \label{eqn4}
\end{equation}

Since $\widetilde{L}_0$ is self-adjoint then $\langle \bar{g} |\widetilde{L}_0 \bar{g} \rangle $ is real, therefore from \eqref{eqn4} it follows
that $\Lambda \braket{\bar{g}}{\bar{f}}$ is also real, i.e.,  $(\Lambda \braket{\bar{g}}{\bar{f}})^\star
=\Lambda \braket{\bar{g}}{\bar{f}}$. From the last condition and after some straightforward calculations 
we deduce that there exists $\gamma\in\mathbb{R}$ such that 
\begin{equation}\label{rel}
\braket{\bar{g} }{\bar{f}} = \gamma {\Lambda}^\star. 
\end{equation}
Now we multiply Eq. \eqref{eqn1b} by  $\bar{f}$ and integrate over $\mathbb{R}$, to obtain
\begin{align}\label{eqn5aa}
\langle \bar{f} | {L}_1 \bar{f} \rangle + \beta \langle \bar{f} | L_1  \psi \rangle &=  - \Lambda \langle \bar{f} | \bar{g} \rangle.
\end{align}
Taking the complex conjugation in the last equation and using relation  \eqref{rel} it transforms into
\begin{equation}\label{eqn6a}
\langle \bar{f} | {L}_1 \bar{f} \rangle + \beta^\star \langle \psi | L_1 \bar{f}  \rangle =  - \gamma \left(\Lambda^\star \right)^2 .
\end{equation}
Next we multiply Eq. \eqref{eqn1b} by $\psi$ and integrate over $\mathbb{R}$ to get
\begin{align}
\label{eqn5a}
\langle \psi | {L}_1 \bar{f}\rangle + \beta\langle \psi |L_1  \psi \rangle&= -\Lambda \alpha \langle \psi |\psi \rangle.
\end{align}
 Subtracting Eq. \eqref{eqn6a} from Eq. \eqref{eqn5a} multiplied by  $\beta^\star$,
and using the fact that $\alpha = -\Lambda \beta/\varepsilon$, we obtain
\begin{equation}\label{eqn7}
\begin{split}
\langle \bar{f} | {L}_1 \bar{f} \rangle - |\beta|^2\langle \psi |L_1  \psi \rangle=   - \gamma \left(\Lambda^\star \right)^2 -|\beta|^2 \Lambda^2  \frac{\langle \psi |\psi \rangle}{\varepsilon}.
\end{split}
\end{equation}
Since $L_1$ is self-adjoint, the left hand side of \eqref{eqn7} is real. 
Therefore,  
\begin{equation}\label{eqn8}
\begin{split}
\Im\left( - \gamma \left(\Lambda^\star \right)^2 -|\beta|^2 \Lambda^2  \frac{\langle \psi |\psi \rangle}{\varepsilon} \right) = 0.
\end{split}
\end{equation}
Now, we notice that when the quadruplet is formed, $\Lambda_i, \Lambda_r \neq 0$, then
\begin{equation} \label{eqn9}
\gamma = |\beta|^2 \frac{\langle \psi |\psi \rangle}{\varepsilon}.
\end{equation}
Substituting Eq. \eqref{eqn9} into Eq. \eqref{eqn7}, yields
\begin{equation}\label{eqn10}
\begin{split}
\langle \bar{f} | {L}_1 \bar{f} \rangle - |\beta|^2\langle \psi |L_1  \psi \rangle =  \left(\Lambda_i^2-\Lambda_r^2\right)|\beta|^2 \frac{\langle \psi |\psi \rangle}{\varepsilon}.
\end{split}
\end{equation}
We will show that, for $\beta\neq0$,  the left-hand side of  \eqref{eqn10} is positive, therefore, $\Lambda_i^2 > \Lambda_r^2$. 
First, from Eq.\ \eqref{psi} follows 
\begin{equation}\label{L_1>0}
\langle \psi |L_1  \psi \rangle 
= -2\kappa(2-\varepsilon) \int_{\mathbb{R}} \psi^{2\kappa+2} \, dx <0.
\end{equation}
Next we prove that $\langle \bar{f} | {L}_1 \bar{f} \rangle \geq0$.
To do that, we use the following result (see Proposition 2.7 of Ref. \cite[pag. 477]{weinstein:1985}):

\medskip
\noindent\textbf{Proposition B1:} $L_1$ verifies for $\kappa < 2$ that
\begin{equation}\label{ineq}
\inf_{\braket{\xi}{\psi} = 0} \braket{\xi}{L_1 \xi} = 0.
\end{equation}
In other words, for $\kappa < 2$, $\braket{\xi}{L_1 \xi}\geq0$ for all $\xi$ orthogonal to 
$\psi$, i.e., $L_1$ is positive definite in the subspace orthogonal to $\psi$. 
Combining this two results it follows that 
$\Lambda_i^2 > \Lambda_r^2$ for any quadruplet obtained when $\kappa< 2$,
and therefore the dotted black line and the dashed red line in Fig.~\ref{fig1} 
are indeed identical for $\kappa< 2$.

Finally, 
 $\beta\neq0$. Indeed, if $\beta=0$,
then, according to Eq. \eqref{eqn10}, 
$\langle \bar{f} | {L}_1 \bar{f} \rangle=0$. Moreover, in this case
$\alpha=0$, $\bar{f}=f_c$ and $\bar{g}=\widetilde{g}_c$.
So the infimum in Eq. \eqref{ineq} is attained when $\xi=f_c$, therefore, for any $h$
$$
0 = \left.\frac{d}{d\alpha} \braket{f_c+\alpha h}{L_1 (f_c+\alpha h)} \right|_{\alpha = 0}
=2\Re \braket{h}{L_1 f_c}.
$$
Thus, $L_1 f_c=0$. 
Using the fact that Eq. \eqref{eqn1b}, in this case, reads
$L_1 f_c=-\Lambda \widetilde{g}_c$, it follows that $\widetilde{g}_c=0$, and
Eq. \eqref{eqn3a} yields $f_c=0$, which is a contradiction.

\subsection*{Acknowledgments}
We thank Ricardo Carretero for sending us a preliminary version of Ref. \cite{carretero:2024} and Igor Bara\-shen\-kov for drawing the Ref. \cite{barashenkov:1990} to our
attention. F.C.N. acknowledges financial support through the Programa de Iniciación a la Investigación PI3 (IMUS) and hospitality at IMUS (University of Seville). 
R.A.N. was partially supported by PID2021-124332NB-C21
(FEDER(EU) Ministerio de Ciencia e Innovación-Agencia Estatal de
Investigación, Spain) and FQM-415 (Junta de Andalucía, Spain). N.R.Q.
was partially supported by the Spanish projects PID2020-113390GB-
I00 (MICIN), and FQM-415 (Junta de Andalucía, Spain).


%
\end{document}